\documentclass{ws-ijns}

\usepackage{amsmath}
\usepackage{amssymb}
\usepackage{hyperref}
\usepackage{algpseudocode}
\usepackage{algorithm}
\usepackage{color,xcolor}
\usepackage{latexsym,graphicx,url,listings}
\usepackage[super,sort,compress]{cite}

\usepackage[utf8]{inputenc}
\usepackage[english]{babel}

\newcommand{\gcomment}[1]{\color{black!50}\Comment{#1}\color{black}}

\begin{document}
	\catchline{34}{7}{2024}{2450038}{}

    \markboth{J. Hern\'andez-Tello et al.}{Sparse Spiking Neural-like Membrane Systems on Graphics Processing Units}

	\title{Sparse Spiking Neural-like Membrane Systems on Graphics Processing Units\footnote{When citing this paper, please use the following: J. Hernández-Tello, M.A. Martínez-del-Amor, D. Orellana-Martín, F.G.C. Cabarle, Sparse Spiking Neural-Like Membrane Systems on Graphics Processing Unit. \textit{International Journal of Neural Systems} 34, 07 (2024), 2450038.  \url{https://doi.org/10.1142/S0129065724500382}}}
	\author{Javier~Hern\'andez-Tello$^1$, Miguel~\'A.~Mart\'inez-del-Amor$^1$, David~Orellana-Mart\'in$^1$, Francis George C. Cabarle$^{1,2}$}
        \address{$^1$Research Group on Natural Computing,\\
	Department of Computer Science and Artificial Intelligence,\\
        I3US, SCORE lab,\\
	Universidad de Sevilla, Avda. Reina Mercedes s/n, 41012, Sevilla, Spain\\
	E-mail: \{jhtello,mdelamor,dorellana,fcabarle\}@us.es\\
 $^2$Dept. of Computer Science, University of the Philippines Diliman, Quezon city, Philippines, 1101\\
 E-mail: fccabarle@up.edu.ph}

	\maketitle
	\begin{abstract}

 The parallel simulation of Spiking Neural P systems is mainly based on a matrix representation, where the graph inherent to the neural model is encoded in an adjacency matrix. The simulation algorithm is based on a matrix-vector multiplication, which is an operation efficiently implemented on parallel devices. However, when the graph of a Spiking Neural P system is not fully connected, the adjacency matrix is sparse and hence, lots of computing resources are wasted in both time and memory domains.  
 For this reason, two compression methods for the matrix representation were proposed in a previous work, but they were not implemented nor parallelized on a simulator. In this paper, they are implemented and parallelized on GPUs as part of a new Spiking Neural P system with delays simulator. Extensive experiments are conducted on high-end GPUs (RTX2080 and A100 80GB), and it is concluded that they outperform other solutions based on state-of-the-art GPU libraries when simulating Spiking Neural P systems. 
	\end{abstract}


\begin{multicols}{2}
\section{Introduction}

Spiking Neural P (SNP) systems \cite{snp} are membrane systems composed of a directed graph, where nodes are neurons that communicate with a singleton alphabet (spike object). SNP systems have been studied widely for computability, complexity and real-life applications \cite{snp-applications,rong2018spiking}. As a result, there is a vast amount of SNP system variants including: delays, division, budding, astrocytes \cite{snp-astro}, weights \cite{snp-weights}, 
dendrites \cite{dendrite},
delays on synapses \cite{delaySynap}, stochastic firing \cite{lazo_starsnp2021}, scheduled synapses \cite{cabarle2017scheduled}, 
extended channels \cite{extendedChannels} and non-linear \cite{nonLinear}. Therefore, the research community has been developing simulators  specific for each variant, given that each of them requires a different semantics to reproduce.

Most sequential simulators of SNP systems \cite{snp} make use of \textit{ad-hoc} representations, specifically defined for a variant \cite{simu-time}, while the parallel simulation of SNP systems \cite{ohips,martinez2015simulating} has been mainly based on a matrix representation \cite{snp-matrix} for the vanilla model. The key concept of this representation is to encode the inherent graph of SNP systems on an adjacency matrix, and a vector-matrix multiplication to perform one computation step. This design can be extended to SNP system variants with more vector definitions and specific algorithms replacing the vector-matrix multiplication, while keeping the algebraic definition and operations as simple as possible \cite{dend-tool,zak_snpsnpmat2019,rssnp_solngpu2019}. 

The family of simulators cuSNP use as a core this matrix representation for their simulation algorithm \cite{cusnp,cusnp-nondet,improvecusnp}, and implement it on Graphics Processing Units (or simply, GPU). GPUs are parallel devices with thousands of parallel cores \cite{cudabook} that have been used to accelerate the simulation of P systems \cite{martinez2015simulating,ohips}. 

However, this matrix representation can be sparse, i.e. having a majority of zero values in the adjacency graph. The main cause is that the directed graph of SNP systems is not usually fully connected. Sparse vector-matrix operations (SpMV) are well known and natural in high performance computing solutions, specially on GPUs \cite{matrixgpu1}. Preliminary works exist in optimising the matrix representation of CuSNP \cite{aboy2019_optcusnp} as well as for WebGL in web browsers\cite{gpusnapse23}. More recently, new compressed matrix representations (named ELL and Optimized) were introduced for several SNP system variants\cite{processes}: standard without delays, with budding and division, and with plasticity. This previous work shows that for SNP systems with dynamic structures, plasticity variant fits better with compressed sparse matrix representations.

The main aims of this work are as follows:  provide GPU-based implementations to these new compression methods for sparse matrix SNP system representation; 
provide extensive experiments to confirm  such methods, using high-end GPUs.
In order to demonstrate that they benefit to GPU-based simulators, they are developed using CUDA, which is the most employed framework for GPU computing today. This work focuses on standard SNP systems, given that the goal is to test and compare the compression methods. Extensions to specific variants should be done for each case in separate works using the design concepts here provided. Nevertheless, an extension to standard SNP systems with delays is given, showing the required changes in the design and the implementation.

Specifically, the contributions of the paper are as follows: (a) the first GPU-parallel simulator for SNP systems with and without delays that uses ELL and Optimized compression methods for the matrix representation; (b) an efficient way to implement simple regular expressions on GPUs, that enables to simulate a wide variety of SNP systems; (c) testing the simulator on four designs: \textit{sparse} (no compression), \textit{ELL}, \textit{Compressed} (previously known as Optimized in \cite{processes}), \textit{cuBLAS}\footnote{\url{https://docs.nvidia.com/cuda/cublas}} and \textit{cuSPARSE}\footnote{\url{https://docs.nvidia.com/cuda/cusparse}}; (d) showing that the simulator using the Compressed design outperforms the rest of alternatives, including the state-of-the-art {\it cuBLAS} and {\it cuSPARSE} on two benchmarks; (e) scalability test to show the maximum instance that a high-end GPU can handle using these designs. The experimental results also show there is room for improvements and more research, since compressed representation of SNP systems will help to better deploy and scale out models. 
On the one hand the implementations and extensive experiments in the present work highlight the value of the previous theory\cite{processes}.
On the other hand, the present work provides value on further theory: how to continue optimising in a fundamental way the matrix representation and simulation, also for other variants, of SN P systems.

The paper is structured as follows: Section \ref{sec:baserep} gives definitions for the matrix representation of SNP systems;  Section \ref{sec:gpu} summarizes fundamental concepts of GPU computing and sparse matrix implementation; Section \ref{sec:spmv-snp} gives a short description of the design of compressed matrix representation of SNP systems; Section \ref{sec:impl} shows the implementation details of the new simulators on GPUs; Section \ref{sec:res} shows the results of the simulators; Section \ref{sec:conclusions} discusses conclusions and future work.

\section{Baseline matrix representation of SNP systems}
\label{sec:baserep}

Let us briefly review the syntactical ingredients of a \emph{Spiking Neural P system with delays} of degree $q \geq 1$. It is a tuple $\Pi = (O, syn, \sigma_1, \dots, \sigma_q, i_{out})$, where $O$ is the singleton alphabet containing the spike symbol $a$; $syn$ represents the arcs of a directed graph by means of a set of pairs $(i,j)$ (i.e. neuron $i$ has a synapse with $j$), and reflexive synapses are not allowed, that is, synapses of the form (i, i);  $\sigma_1, \dots, \sigma_q$ are {\it neurons} of the form $\sigma_i = (n_i, R_i), 1 \leq i \leq q$, where $n_i$ is the initial number of spikes in the neuron, and $R_i$ is the finite set of rules associated to the neuron; $i_{out}$ is the label of the output neuron. Let us define $m$ as the total amount of rules defined in $\Pi$:  $m=\sum_{i=1}^q |R_i|$ \cite{processes}. Finally, The rules can be of two types: $E/a^c \rightarrow a^p; d$ (firing rule), with $E$ being a {\it regular expression} over $\{a\}$, $c \geq p \geq 1$, and $d$ the delay saying for how many steps the neuron is closed; or $a^s \rightarrow \lambda$ (forgetting rule), for some $s \geq 1$. More details about such expressions,  syntax and semantics of SNP systems, are in \cite{snp,cusnp-nondet,appsnp}, a recent survey\cite{snpsurv_naco2022} and the handbook\cite{mem-handb}. %This includes regular expressions, what constitutes a valid spiking vector. 

Next, it is described the matrix representation for SNP systems with delays in order to enable a linear-algebra-based simulator. More details can be found in \cite{snp-matrix,cusnp,processes}. Many extensions have followed such as supporting non-determinism \cite{cusnp-nondet}, but they are not covered in this work. The general matrix representation of standard SNP systems consists of mainly three structures: a \textit{Configuration Vector} $C_k$ (saying for each transition step $k$, the amount of spikes in each neuron), a \textit{Spiking Vector} $S_k$ (saying for each step $k$ if a rule is going to fire), and a Spiking Transition Matrix $M_{\Pi}$ (saying how each rule affects each neuron). Their definitions are developed below, but this first view is enough to define how the configuration of a SNP system transits from step $k-1$ to $k$ by simply: $C_k = S_k \cdot M_\Pi + C_{k-1}$ \cite{snp-matrix}.

\begin{algorithm}[H]
	\caption{MAIN PROCEDURE: simulating one computation of a SNP system with delays}
	\label{alg:snpgeneral}
	{\small \begin{algorithmic}[1]
			\Require A SNP system $\Pi$ of degree $q$ with $m$ rules, and a max number $L>0$ of computational steps.
			\Ensure A computation of the system.
			\State $(C_0,M_\Pi,RV_{\Pi},N_{\Pi})\ \gets$ \Call{INIT}{${\Pi}$}
			\State $D_0\ \gets\ \{0,\ldots,0\}$
			\State $k\ \gets\ 0$
			\Repeat 
			\State $SV_k\ \gets\ $\Call{SV\_CALC}{$C_{k},D_k,RV_{\Pi},N_{\Pi}$} 
			\If{$SV_k \neq \emptyset$}
			\State $S_k\  \gets\ $\Call{SAMPLE\_ONE}{$SV_{k}$}	
			\State $C_{k+1}\ \gets\ $ \Call{STEP}{$C_{k},S_K,D_k,M_\Pi,RV_{\Pi}$}    
			\State $D_{k+1}\ \gets\ $\Call{UPDATE\_DELAYS}{$D_k,S_k$}
			\State $k\ \gets\ k+1$
			\EndIf
			\Until{$k = L \vee (SV_k = \emptyset \wedge D_k = \{0,\ldots,0\})$}          
			\State \Return $C_0 \ldots\ C_{k-1}$   
	\end{algorithmic}}
\end{algorithm}

Next, our matrix representation employed in parallel simulators for a SNP system with delays is presented. Let us assume a SNP system of degree $q$ with $m$ rules. Moreover, for the sake of simplicity, current representation only allows the following three types of regular expressions in the rules: 

	\begin{itemize}
		\item Type one ($e^*$). Rule can be activated at any possible condition.
		\item Type two ($e^+$). Rule may activate if its corresponding neuron contains at least 1 spike.
		\item Type three ($e^n$). Rule may activate if and only if its neuron contains exactly n spikes.		
	\end{itemize}

Algorithm \ref{alg:snpgeneral} is the  pseudocode to perform one computation of a SNP system based on the matrix representation, and the following data structures:
\begin{itemize}
    \item {\it Spiking vector} $S_k$ of length $m$, stores which rules are active at each computation step.
    \item {\it Set of Spiking Vectors} $SV_k$, which stores all spiking vectors that can be computed in a computation step $k$.
    \item {\it Rule Vector} $RV_{\Pi}$, of size $m$. It contains the information for each rule in the model. Each item contains the following data:
        \begin{itemize}
            \item The regular expression, which is composed of two elements:
            \begin{itemize}
                \item The multiplicity ($E_n$). It indicates the minimum or the exact multiplicity of required spikes: 0 for type one, 1 for type two and $n$ for type three.
                \item The type of regular expression ($E_i$). It indicates if the expression is a minimum or an exact operation. 0 for types one and two, 1 for type three.
            \end{itemize}
            \item The number of spikes $c$ that are consumed in the neuron containing the rule.
            \item The number of spikes $p$ sent. This element is skipped when the transition matrix contains it, i.e., for sparse and ELL formats (more in Section \ref{sec:spmv-snp}).
            \item The delay $d$ that is applied to the neuron when the rule is applied.
            \item The id of the neuron ($nid$) that contains it.
        \end{itemize}
    \item {\it Neuron-Rule Map Vector} $N_{\Pi}$, which, as the name implies, maps each neuron $\sigma_i$ to its rule set $R_i$. $N_{\Pi}[i]$ is the index of the first rule in that set and $N_{\Pi}[i+1]-1$ is the last one. The vector contains $q+1$ elements. It is initialized by adding to the index stored in $N_{\Pi}[i-1]$ the total number of rules stored in the neuron $\sigma_{i-1}$ (i.e $N_\Pi[i]=N_{\Pi}[i-1]+|R_{i-1}|$, for  $2\leq i\leq q+1$). 
    \item {\it Configuration vector} $C_k$, of size $q$. It stores, for each transition step $k$, the number of spikes available to each neuron. $C_0$ corresponds to the initial configuration; i.e., $C_0[i]=n_i$ for $\sigma_i=(n_i, R_i)$.
    \item {\it Transition matrix} $M_{\Pi}$, of size $q\times m$, in which information about the synapses and rules of the model is stored.
    \item {\it Delays Vector} $D_k$, with $q$ elements (one per neuron), indicating the state (open or closed) of each neuron and, if closed, how many transition steps before it to reopens. 
\end{itemize}

At the beginning of the computation it will be necessary to initialize ({\tt INIT} function) the \textit{Configuration vector} ($C_0$), the \textit{Transition Matrix} ($M_\Pi$), Rule Vector ($RV_{\Pi}$) and Neuron-Rule Map Vector ($N_{\Pi}$) with the initial data of the system $\Pi$. Moreover, the first Delays Vector ($D_0$) is initialized with zeroes. Once this is done, the main simulation loop starts. First, it will be necessary to calculate the set of all possible \textit{Spiking Vectors} ($SV_k$) with the {\tt SV\_CALC} function, by using the rules information ($RV_\Pi$ and $N_\Pi$) and the current configuration of the system ($C_k$ and $D_k$). From this set, a spiking vector ($S_k$) will be randomly selected with the {\tt SAMPLE\_ONE} function, in order to simulate nondeterminism in the computation. $S_k$ will subsequently be used, together with the transition matrix ($M_\Pi$) and the rule vector ($RV_\Pi$), to compute the next configuration vector ($C_{k+1}$). This computation is carried on with the {\tt STEP} function, and can be implemented with a single Matrix-Vector operation, that is, $C_{k+1} = S_k \cdot M_\Pi + C_{k}$. Once this step is completed, the delays vector ($D_{k+1}$) gets updated with {\tt UPDATE\_DELAYS}, decreasing the delay counter for each closed neuron to indicate that a computation step has just been completed, or otherwise the delay is increased with the one of the rule executed (according to the Spiking Vector $S_k$). The simulation will end either when a maximum number of steps ($L$) has been reached, or when no rule can be applied in the current configuration (the spiking vector is empty). In the case that the model makes use of delays, it will also be necessary to check that all neurons are open (the Delays Vector has only null values), in case some rules are pending to be executed. The list of all configuration vectors $C_k$ is returned as output of the simulator, which corresponds to a computation (maybe truncated) of the SNP system.

\section{GPU computing and sparse matrices}
\label{sec:gpu}

Sparse matrix vector multiplication (SpMV) is a widely and critical operation in many scientific fields, including deep learning (e.g. sparse convolutions) and graph analytic (e.g. page rank). Sparse matrices can facilitate applications scaling in memory, since the growth can be superlinear. Thus, their compression is cornerstone for these applications, and it has been widely studied in the literature \cite{sparse-survey-16,sparse-survey}. 

GPUs have been settled as a powerful technology for High Performance Computing, driving successful areas today such as machine learning. CUDA is the main programming model and language for GPUs, although it is only supported for NVIDIA brand. The key element of CUDA is the \textbf{kernel}, which is a function that gets executed on the GPU. This execution is based on parallel threads that run, each one, the same code of the kernel.

GPUs devices are tailored for data parallelism, and hence, they are good at linear algebra operations. In this sense, cuBLAS library is a CUDA implementation of the BLAS (Basic Linear Algebra Subprograms) subroutines. It is used especially in the acceleration of High Performance Computing (HPC) and Artificial Intelligence (AI) applications. This library is already included in the CUDA ecosystem.

Moreover, SpMV has been also extensively studied for GPUs, and several compression formats have been defined that fit well for data parallelism \cite{matrixgpu1,cudabook}. Next, the two formats that have been employed in this work are summarized:. 
\begin{itemize}
    \item \textit{CSR} uses for the representation a vector containing only the non-zero values, another vector of the same size to indicate the column of each value, and a third vector with as many elements as rows, that indicates the beginning of each row in the two previous vectors. In this format, the random access to the elements is driven by rows. 
    \item \textit{ELL} first calculates the transpose of the matrix, which improves the data coalescing in GPUs (a critical memory access pattern in these devices). The compressed representation of the matrix is another where the number of columns is equal to the original number rows, and the number of rows is the maximum number of non-null values that can exist in any row in the original matrix. Each element  contains a pair of elements containing the column to which a non-null element belongs to, and the corresponding value. The memory required for this format is larger than for CSR, since it has null elements (but at a slower scale than in the original matrix). On the contrary, algorithms are more efficient and easier to implement.
\end{itemize}

CUDA also contains a library to handle sparse matrices, named cuSPARSE. It implements numerous subroutines for performing various algebraic operations with sparse matrices. The library is recommended for use on matrices and vectors where the number of null elements is more than 95\% of the total number of elements. cuSPARSE assumes that input and output reside in device (GPU) memory. 

\section{Compression of sparse Spiking Transition Matrices}
\label{sec:spmv-snp}

In order to achieve a good performance on GPUs, the \textit{Rule Vector} $RV_{\Pi}$ is implemented using a CSR-like format \cite{cudabook,matrixgpu1}, so that rules of the form $E/a^c \rightarrow a^p$ (also forgetting rules are included, assuming $p=0$ and $E = a^c$) can be represented by three arrays that store the regular expression associated into the rule, and the values $c$ and $p$. In order to go from a given neuron to its set of rules, it is enough to access the Neuron-Rule map vector $N_{\Pi}$, as mentioned above. For all compression formats discussed next, both $RV_{\Pi}$ and $N_{\Pi}$ are required in order to select the rules and compute a spiking vector.

$M_\Pi$ is the data structure that constitutes the bottleneck in terms of memory and performance, since its size depends on both the number of neurons $q$ and the number of rules $m$. However, it can be very sparse, that is, with a majority of zero values. $M_\Pi$ contains the adjacency matrix of the graph structure of a SNP system. Usually, this graph is not fully connected, but each neuron is connected to a limited number of other neurons. For instance, the transition matrix for the SNP systems without delays sorting natural numbers \cite{appsnp} contains 75\% of zeroes, as  seen in the example shown in Table \ref{table:exampleSparse}.

\begin{tablehere}
%\resizebox{\textwidth}{!}{%
\tbl{Example of a sparse Transition Matrix. This corresponds to the SNP system for sorting 3 natural numbers, as shown in Figure 3 of Ionescu \& Sburlan \cite{appsnp}. Columns show each neuron, and rows the rules. Each row is labelled by an unique rule identifier and the neuron where it belongs. The numbering of the rules correspond from top to bottom as shown in the mentioned figure; e.g. rule $r_4$ is $a^3 \rightarrow a; 0$ (inside neuron $s_1$), rule $r_5$ is $a^2 \rightarrow \lambda; 0$ (in $s_1$), rule $r_6$ is $a \rightarrow \lambda $ (in $s_1$). It contains a total of 108 elements.
\label{table:exampleSparse}	}
{\begin{tabular}{|l||r|r|r|r|r|r|r|r|r|}
            \hline
             & {\textbf{$i_1$}} & {\textbf{$i_2$}} & {\textbf{$i_3$}} & {\textbf{$s_1$}} & {\textbf{$s_2$}} & {\textbf{$s_3$}} & {\textbf{$o_1$}} & {\textbf{$o_2$}} & {\textbf{$o_3$}} \\ \hline \hline
\textbf{$r_1/i_1$}  & \textbf{-1}                          & 0                                    & 0                                    & \textbf{1}                           & \textbf{1}                           & \textbf{1}                           & 0                                    & 0                                    & 0                                    \\ \hline
\textbf{$r_2/i_2$}  & 0                                    & \textbf{-1}                          & 0                           & \textbf{1}                           & \textbf{1}                           & \textbf{1}                           & 0                                    & 0                                    & 0                                    \\ \hline
\textbf{$r_3/i_3$}  & 0                                    & 0                                    & \textbf{-1}                          & \textbf{1}                           & \textbf{1}                           & \textbf{1}                           & 0                                    & 0                                    & 0                                    \\ \hline
\textbf{$r_4/s_1$}  & 0                                    & 0                                    & 0                                    & \textbf{-3}                          & 0                                    & 0                                    & \textbf{1}                           & \textbf{1}                           & \textbf{1}                           \\ \hline
\textbf{$r_5/s_1$}  & 0                                    & 0                                    & 0                                    & \textbf{-2}                          & 0                                    & 0                                    & 0                                    & 0                                    & 0                                    \\ \hline
\textbf{$r_6/s_1$}  & 0                                    & 0                                    & 0                                    & \textbf{-1}                          & 0                                    & 0                                    & 0                                    & 0                                    & 0                                    \\ \hline
\textbf{$r_7/s_2$}  & 0                                    & 0                                    & 0                                    & 0                                    & \textbf{-2}                          & 0                                    & 0                                    & \textbf{1}                           & \textbf{1}                           \\ \hline
\textbf{$r_8/s_2$}  & 0                                    & 0                                    & 0                                    & 0                                    & \textbf{-3}                          & 0                                    & 0                                    & 0                                    & 0                                    \\ \hline
\textbf{$r_9/s_2$}  & 0                                    & 0                                    & 0                                    & 0                                    & \textbf{-1}                          & 0                                    & 0                                    & 0                                    & 0                                    \\ \hline
\textbf{$r_{10}/s_3$} & 0                                    & 0                                    & 0                                    & 0                                    & 0                                    & \textbf{-1}                          & 0                                    & 0                                    & \textbf{1}                           \\ \hline
\textbf{$r_{11}/s_3$} & 0                                    & 0                                    & 0                                    & 0                                    & 0                                    & \textbf{-2}                          & 0                                    & 0                                    & 0                                    \\ \hline
\textbf{$r_{12}/s_3$} & 0                                    & 0                                    & 0                                    & 0                                    & 0                                    & \textbf{-3}                          & 0                                    & 0                                    & 0     \\ \hline                             
\end{tabular}}%
%}
\end{tablehere}

This compression reduces the memory footprint of the simulators and hence, gain acceleration. In \cite{processes}, three variants to implement $M_\Pi$ are proposed:

\begin{itemize}
	\item \textit{Sparse}: this implementation has no compression, as defined as above. See Table \ref{table:exampleSparse} for an example.  The $+p$ value is not stored in $RV_{\Pi}$ since it is not required for selecting a rule.
	\item \textit{ELL}: this implementation is based on the ELL compression, as explained above (See example in Table \ref{table:exampleELL}), where:
 	
	\begin{itemize}
		\item The transition matrix is now $M^s_\Pi$. The number of rows is the maximum amount of non-zero values in a row of $M^s_\Pi$, denoted by $z$. It can be shown that $z=MaxOutDegree + 1$, where $MaxOutDegree$ is the maximum output degree of the nodes in the graph $syn$ (i.e. the out degree in the neurons of the SNP system). In general, a column devoted for a rule $E/a^c \rightarrow a^p$ contains values $+p$ for every neuron connected with the source neuron (i.e. where it belongs to), and a value $-c$ for consuming the spikes in that source neuron.
		\item The values inside columns can be sorted, so that the consumption of spikes ($-c$ values) are placed at the first row. In this way, all threads can start with the same task, consuming spikes. Moreover, the loop along the columns can be ended prematurely, once 0 values are encountered.
		\item Every position is a pair where the first element is a neuron label, and the second is the amount of spikes ($+p$ or $-c$).
	\end{itemize}
	\item \textit{Compressed} (a.k.a. \textit{Optimized}): the transition matrix can be split in order to avoid, for each rule, replicating the generation of spikes ($+p$) for all synapses. In fact, the amount of spikes to consume ($-c$) is already present in the Rule Vector $RV_{\Pi}$, so including this information again in the transition matrix is redundant. It is only required to add the $+p$ value in $RV_{\Pi}$. Thus, only the following modifications are needed:
 %the synapses can be stored separately into a new matrix, and with the following modifications:
	
	\begin{itemize}
	  \item \textit{Synapse matrix}, $Sy_\Pi$, which replaces $M_{\Pi}$. It has a column per neuron $i$, and a row for every neuron $j$ such that $(i,j) \in Syn$ (there is a synapse).	That is, every element of the matrix corresponds to a synapse or null. The latter is necessary given that the number of rows equals to the maximum output degree in the neurons of the SNP system and padding is required.
       			
		\item The \textit{Spiking vector} gets smaller, containing only $q$ positions, one per neuron, and stating which rule $0\leq r \leq m$ is selected.
        \item  The Rule Vector $RV_{\Pi}$ contains the $+p$ value for each rule, which is 0 for forgetting rules.
	\end{itemize}	
\end{itemize}

\begin{table*}

\tbl{Example of a ELL Transition Matrix. This corresponds to the SNP system for sorting 3 natural numbers, as shown in Figure 3 of Ionescu \& Sburlan \cite{appsnp}. Columns show each rule, and rows the non-null values. Each pair indicates a value and the corresponding neuron. It contains a total of 60 positions but only 120 values (considering the pairs).
\label{table:exampleELL}	}
{
\begin{tabular}{|c|c|c|c|c|c|c|c|c|c|c|c|}
            \hline
             {\textbf{$r_1$}} & {\textbf{$r_2$}} & {\textbf{$r_3$}} & {\textbf{$r_4$}} & {\textbf{$r_5$}} & {\textbf{$r_6$}} & {\textbf{$r_7$}} & {\textbf{$r_8$}} & {\textbf{$r_9$}} & {\textbf{$r_{10}$}} & {\textbf{$r_{11}$}} & {\textbf{$r_{12}$}} \\ \hline \hline

(-1,$i_1$)         & (-1,$i_2$)         & (-1,$i_3$)         & (-3,$s_1$)         & (-2,$s_1$)     & (-1,$s_1$)     & (-2,$s_2$)         & (-3,$s_2$)     & (-1,$s_2$)     & (-1,$s_3$)      & (-2,$s_3$)      & (-3,$s_3$)      \\\hline
{(1,$s_1$)} & {(1,$s_1$)} & {(1,$s_1$)} & {(1,$o_1$)} &             &             & {(1,$o_2$)} &             &             &        (1,$o_3$)       &              &              \\\hline
{(1,$s_2$)} & {(1,$s_2$)} & {(1,$s_2$)} & {(1,$o_2$)} &             &             & {(1,$o_3$)} &             &             &              &              &              \\\hline
{(1,$s_3$)} & {(1,$s_3$)} & {(1,$s_3$)} & {(1,$o_3$)} &             &             &                 &             &             &             &              &            \\\hline    
\end{tabular}}%
%}
\end{table*}

\begin{tablehere}
%\resizebox{\textwidth}{!}{%
\tbl{Example of Compressed Synapse Matrix. This corresponds to the SNP system for sorting 3 natural numbers, as shown in Figure 3 of Ionescu \& Sburlan \cite{appsnp}. Columns show each neuron, and rows the neurons to which each one is connected. It contains 27 values.
\label{table:exampleCompr}	}
{
\begin{tabular}{|c|c|c|c|c|c|c|c|c|}
            \hline
$i_1$ & $i_2$ & $i_3$ & $s_1$ & $s_2$ & $s_3$ & $o_1$ & $o_2$ & $o_3$ \\\hline \hline
$s_1$ & $s_1$ & $s_1$ & $o_1$ & $o_2$ & $o_3$ &      &      &      \\\hline
$s_2$ & $s_2$ & $s_2$ & $o_2$ & $o_3$ &      &      &      &      \\\hline
$s_3$ & $s_3$ & $s_3$ & $o_3$ &      &      &      &      &      \\\hline
\end{tabular}}%
%}
\end{tablehere}

For more information and pseudocode of each algorithm, the reader is referred to \cite{processes}.

\section{Implementation}
\label{sec:impl}

The current state of the source code is available at \url{https://github.com/RGNC/sparse_snp}. In what follows, the stages of the simulator are depicted. This design is based on a C++ object that contains the model, which can be configured using a customized API for SNP systems.

\subsection{Initialization}

The target model can be defined in the C++ object by providing the following information:

\begin{enumerate}
	\item The initial configuration of the model; i.e. the spikes each neuron will have in the beginning. 
	
	\item The rules for each neuron. 
 The number of spikes to be consumed, the spikes to be produced, a regular expression, and a delay are provided. 
	
	\item Add synapses of the SNP system. Once the initial configuration is set and all the rules of the model have been specified, the final step in the creation of the model is adding synapses. This information will be used to create the transition matrix, a vital part for the computation of the model. The simulator first initializes a sparse matrix, which is compressed afterwards for the corresponding implementation. This is done in this way to make the definition of the SNP system flexible for the user, otherwise the simulator would require parameters such as $z$ beforehand to work with compressed representations only.
	
\end{enumerate}

\subsection{Simulation Loop}

The parallel simulator implements Algorithm \ref{alg:snpgeneral} with CUDA kernels as follows:

\begin{itemize}
    \item \textbf{{\tt SV\_CALC}}: The Spiking Vector Calculation is implemented with a CUDA kernel by launching as many threads as neurons in the model. The purpose of thread $i$ will be to examine the set of rules $R_i$ (i.e. rules of neuron $i$) and determine which rules can be enabled. Only one will get selected randomly for each neuron; however, in this exploratory version, a simple deterministic loop traverses the rules of each neuron, and only the first applicable rule is chosen. Therefore, this function only returns one Spiking Vector, so it is merged with {\tt SAMPLE\_ONE}.
    A rule is applicable if its regular expression fits with the number of spikes ($n$) in the neuron. Since only three types of regular expressions are allowed, this check can be performed efficiently by the threads without divergence, since it consists in only one Boolean operation: $((E_i = 0) \wedge (n \geq E_n)) \vee ((E_i = 1) \wedge (E_n = n))$. The pseudocode is shown in Algorithm \ref{alg:calcspikingvec}. Only for Compressed version, the pseudocode will vary on the position where to write the result in the Spiking Vector $S_k$, since it will contain a position per neuron instead of per rule.
   
    \item \textbf{{\tt STEP}}: To perform a transition step (and obtain a new configuration $C_k$) it will be sufficient to multiply the {transition matrix} $M_{\Pi}$ and the spiking vector $S_k$, as mentioned before, but taking into account the delays. This operation will be implemented on the GPU in several versions, depending on the format in which the transition matrix is compressed. Each version is depicted in the next section.
    
    \item \textbf{{\tt UPDATE\_DELAYS}}: The delays vector contains a counter for each neuron in the system.  These counters indicate the number of steps that need to be taken for the closed neurons to reopen. Once a transition step has been executed, it is necessary to decrement by one all counters greater than 0. This is done in parallel as a CUDA kernel assigning a thread to each position of $D_k$.
        
    \item \textbf{Check stop condition}: The condition is checked on the CPU since the loop is performed by the host.
\end{itemize}

\subsection{Transition Implementations}

The transition matrix has been implemented in five different ways, as explained next.

\subsubsection{Sparse}

In the sparse version, the transition matrix is not compressed. A kernel is launched with as many threads as neurons, which will calculate the new spike value of their assigned neuron. To do this, each thread traverses the column of the transition matrix corresponding to its assigned neuron, and add the values for only the rows corresponding to active rules. The final value is added (or subtracted, as appropriate) to the existing spikes of the neuron (stored in the configuration vector $C_k$).

Also note that, if a neuron is closed due to a delay, it cannot receive or send spikes to other neurons. To check that it is open and can therefore receive spikes, the \textit{delays vector} can be accessed with the neuron index (column index) and see that it contains a value equal to 0. Otherwise, all the values of the column are discarded. In the same way, it will be necessary to check that the values of the column are spikes sent by an open neuron.   

Algorithm \ref{alg:calctrans_sp} shows the pseudocode of the kernel. In the algorithm, {\tt SYNCTHREADS} is the CUDA synchronization barrier of threads (all threads wait until all of them reach that point).

\begin{algorithm}[H]
	\caption{Kernel for calculating the spiking vector.}
	\label{alg:calcspikingvec}
	{\small \begin{algorithmic}[1]
			\Statex
			\Procedure{SV\_CALC}{$C_{k},D_k,RV_{\Pi},N_{\Pi}$} 
			\State $nid\ \gets\ thread\_idx$ \gcomment{A thread per neuron}
			\If{$D_k[nid]=0$} \gcomment{If the neuron is open}
			\For{ $r\ \gets\ N_{\Pi}[nid] \ldots N_{\Pi}[nid+1]-1$}\\ 
			\gcomment{Each rule of $R_{nid}$}
			
			\State $e_n\ \gets RV.En[r]$ \gcomment{Mult of regex}
			\State $e_i\ \gets RV.Ei[r]$ \gcomment{Type of regex}
			\State $n\ \gets\ C_k[nid]$ \gcomment{Mult of neuron}
			\State $regtype_{1,2}\ \gets\ (e_i = 0) \wedge (n \geq e_n)$ 
			\State $regtype_{3}\ \gets\ (e_i = 1) \wedge (e_n = n)$
			
			\If{$regtype_{1,2} \vee regtype_{3}$}\\
			\gcomment{Check if any regex type (1,2,3) fits}
			\State $S_{k}[r] \gets 1$ 
			\gcomment{Update the spiking vector}                
			
			\State Break the loop
			\gcomment{A rule was selected}
			\EndIf
			\EndFor
			\EndIf
			\State \Return $S_k$
			\EndProcedure
	\end{algorithmic}}
\end{algorithm}

\begin{algorithm}[H]
	\caption{Kernel for transition step with Sparse format.}
	\label{alg:calctrans_sp}
	{\small \begin{algorithmic}[1]
			\Statex
			\Procedure{STEP}{$C_{k},S_k,D_k,M_\Pi,RV_\Pi$}
			\State $nid\ \gets\ thread\_idx$ \gcomment{A thread per neuron}
			\If{$D_k[nid]=0$}
			\gcomment{If the neuron is open}
			\For{ $i\ \gets\ 1\ldots m$} 
			
			\State $nidD\ \gets RV.nid[i]$ \gcomment{Destination neuron}
			\If{$D_k[nidD]=0$}
			\gcomment{If neuron is open}
			\State $C_{k+1}[nid] \gets C_k[nid]+S_k[i]\cdot M_\Pi[i,nid]$ 
			\gcomment{Update the configuration vector}                
			
			\State SYNCTHREADS \gcomment{Thread barrier}
			
			\State $S_{k+1}[i]\ \gets\ 0$
			\gcomment{Deactivate rule}
			\EndIf
			\EndFor
			\EndIf
			\State \Return $C_{k+1}$
			\EndProcedure
	\end{algorithmic}}
\end{algorithm}

\begin{algorithm}[H]
	\caption{Kernel for transition step with ELL format.}
	\label{alg:calctrans_ell}
	{\small \begin{algorithmic}[1]
			\Statex
			\Procedure{STEP}{$C_{k},S_k,D_k,M^s_\Pi,RV_\Pi$}
			\State $rid\ \gets\ thread\_idx$ \gcomment{A thread per rule}
			\State $z'\ \gets\ numRows(M^s_\Pi)$ \gcomment{Max out degree plus 1}
			\If{$D_k[RV.nid[rid]]=0$ AND $S\_k[rid]=1$}  \\
			\gcomment{Neuron is open and rule active}
			
			\State $i\ \gets\ 0$
			\Repeat
			\gcomment{Iterate rows}
			
			\State $(nid,x)\ \gets\ M^s_\Pi[i,rid]$ \gcomment{(neuron,spikes)}
			\If{$D_k[nid]=0$}
			\State $\Call{ATOMICADD}{C_{k+1}[nid],x}$ \\
			\gcomment{Update spikes using safe addition}
			\EndIf
			\State $i\ \gets\ i+1$
			\Until{$(nid,x)=null\vee i>z'$}
			
			\State $S_{k+1}[rid]\ \gets\ 0$
			\gcomment{Deactivate rule}
			\EndIf
			\State \Return $C_{k+1}$
			\EndProcedure
	\end{algorithmic}}
\end{algorithm}

\subsubsection{ELL}

For the ELL format of the transition matrix, a kernel is launched with as many threads as rules in the model. Each one deals with a rule in the model, which will initially have to check whether it is active (by checking its state in the \textit{spiking vector}), and if so access its column values to see which neurons come into play and update their respective spikes in the \textit{configuration vector} $C_k$. Note that there may be several rules (columns of the matrix) that send spikes to the same neuron. Therefore, the use of CUDA atomic operations (ATOMICADD) will be necessary when updating the configuration vector. It allows to perform additions safely when threads write over the same value.
A pseudocode is included in Algorithm \ref{alg:calctrans_ell} to facilitate the understanding of the procedure.

Some aspects considered with the ELL version:

\begin{itemize}
    \item The values of all columns are ordered in such a way that the pairs relative to the consumption of spikes are arranged in the first row. In this way, all threads consume spikes at the same time.
    \item The transposition of the matrix favors execution on SIMD processors, such as GPUs.
    \item A negative aspect of using the ELL format is that, due to the design of its structure, the {\it branch divergence} problem in CUDA arises. Recall that a neuron can contain several rules, but can only apply one per transition. This means that there will be several consecutive threads that will be inactive.        
\end{itemize}

    \begin{algorithm}[H]
	\caption{Kernel for transition step with Compressed format.}
	\label{alg:calctrans_opt}
	{\small \begin{algorithmic}[1]
        \Statex
        \Procedure{STEP}{$C_{k},S_k,D_k,Sy_\Pi,RV_\Pi$}
        \State $nid\ \gets\ thread\_idx$ \gcomment{A thread per neuron}
        \State $z\ \gets\ numRows(Sy_\Pi)$ \gcomment{Max out degree}
        
            \If{$D_k[nid]=0$ AND $S_k[nid]> 0$}
            
            \gcomment{Neuron is open and has an active rule}
                \State $rid \gets\ S_k[nid]$
                \gcomment{Get active rule}
                
                \State $(c,p)\ \gets\ (RV_\Pi.C[rid],RV_\Pi.P[rid])$\\
                \gcomment{Obtain pair (c,p) of rule}
                \State $C_{k+1}[nid] \gets\ C_{k+1}-c $
                \gcomment{Consume spikes}
                \State $S_{k+1}[rid]\ \gets\ 0$
                \gcomment{Deactivate rule}
                \State $i\ \gets\ 0$
                \Repeat
                \gcomment{Iterate each row}
                
                    \State $nid2 \gets\ Sy_\Pi[i,nid]$ \gcomment{Neuron rec. spikes}
                    \If{$D_k[nid2]=0$}
                    \gcomment{Neuron is open}
                        \State $\Call{ATOMICADD}{C_{k+1}[nid2],p}$ \\
                        \gcomment{Update spikes with safe addition}
                    \EndIf
                    \State $i\ \gets\ i+1$
                \Until{$Sy_\Pi[i]=null\vee i>z$}

            \EndIf
            \State \Return $C_{k+1}$
        \EndProcedure
	\end{algorithmic}}
    \end{algorithm}

\subsubsection{Compressed}
   
Each thread will be in charge of one neuron $\sigma_i$ (column of the $Sy_{\Pi}$ matrix). The spiking vector is accessed at position $i$, thus obtaining the index of the active rule in $\sigma_i$. The rule vector is accessed with this index, providing its $p$ and $-c$ values. Once these two values are obtained, the spikes of $\sigma_i$ and of the neurons with which it connects (i.e. $j=pres(\sigma_i)$ are modified, which will be those that appear in the columns of the matrix $Sy_{\Pi}$). This same procedure is described in the algorithm \ref{alg:calctrans_opt}.

 \subsubsection{cuBLAS and cuSPARSE}

cuBLAS and cuSPARSE were used to implement the transition matrix and the computation of the next configuration vector. cuBLAS requires no compression, while cuSPARSE compresses the transition matrix in CSR format. The computation of the next configuration vector is performed by a vector-matrix multiplication operation implemented in these libraries. Therefore, due to restrictions in the usage of these libraries, SNP systems without delays are only supported when using these two solutions. The objective is to have a reference framework in order to test the proposed sparse compression formats for SNP systems simulation, since cuBLAS and cuSPARSE are well optimized libraries by experts from NVIDIA.

\section{Experimental results}
\label{sec:res}

The employed GPUs for the experiments were an RTX2080 (2944 cores, 8 GBytes GDDR5, Turing architecture) and an A100 (6912 cores, 80 GBytes HBM2e, Ampere architecture). The simulator was compiled for these two architectures with all optimizations enabled. Two solutions based on SNP systems are selected for the tests:

\begin{itemize}
    \item A family of SNP systems without delays for sorting natural numbers \cite{appsnp}. The examples are built with a worst-case input, where $n$ numbers in inverse order are given. Let us assume that the instance consists of $n=100$ numbers to be sorted (i.e. the sequence is $100,99,\ldots,1$). This instance of the problem would require $q=3n=300$ neurons, $m=n+n^2=10,100$ rules and $z=n=100$ maximum out degree. According to \cite{processes}, the size order would be:
    \begin{itemize}
        \item[*] $m \times q + 3m + 2q + 1 = 3n^3 + 6n^2 + 5n + 1=3,060,501$ for sparse representation;
        \item[*] $m(2z + 5) + 2q + 1 = 2n^3 + 7n^2 + 11n + 1=2,071,101$ for ELL;
        \item[*] $q(z + 3) + 4m + 1 = 7n^2 + 13n + 1=71,301$ for Compressed;
    \end{itemize}
    
    \item A family of SNP systems with delays solving the subset sum problem \cite{uniformsnp}, where the input consists of $V$, that is the set of numbers, and $S$, that is is the objective sum to achieve. This is a non-uniform solution where a non-uniform solution usually means that the size of the system increases (e.g. number of neurons, synapses) as the size of the input increases. In fact, the size of the system depends on the specific set of numbers $V$ provided as input. In the experiments, $n=|V|$, the numbers $v_i$ are randomly generated in the range $[0,50]$, and $S$ is computed selecting only a 20\% of the generated numbers. The simulator only reproduces one path in the computation tree generated by the theoretical non-determinism. In order to illustrate the memory footprint, let us assume that $n=|V|=100$, and for the sake of simplicity for this example, $V=\{1\ldots100\}$. In such a case, the instance will require $q=\sum_{i=1}^{n} v_i + 2n + 2 = 5,252$ neurons, $m=\sum_{i=1}^{n} v_i + 4n + 2 = 5,452$ rules and $z=n=100$ maximum out degree. The size order would be:
        \begin{itemize}
        \item[*] $m \times q + 3m + 2q + 1 = 28,660,765$ for sparse representation;
        \item[*] $m(2z + 5) + 2q + 1=1,128,165$ for ELL;
        \item[*] $q(z + 3) + 4m + 1=562,765$ for Compressed;
    \end{itemize}
\end{itemize}

\begin{figure}[H]
		\centering
		\includegraphics[width=\linewidth]{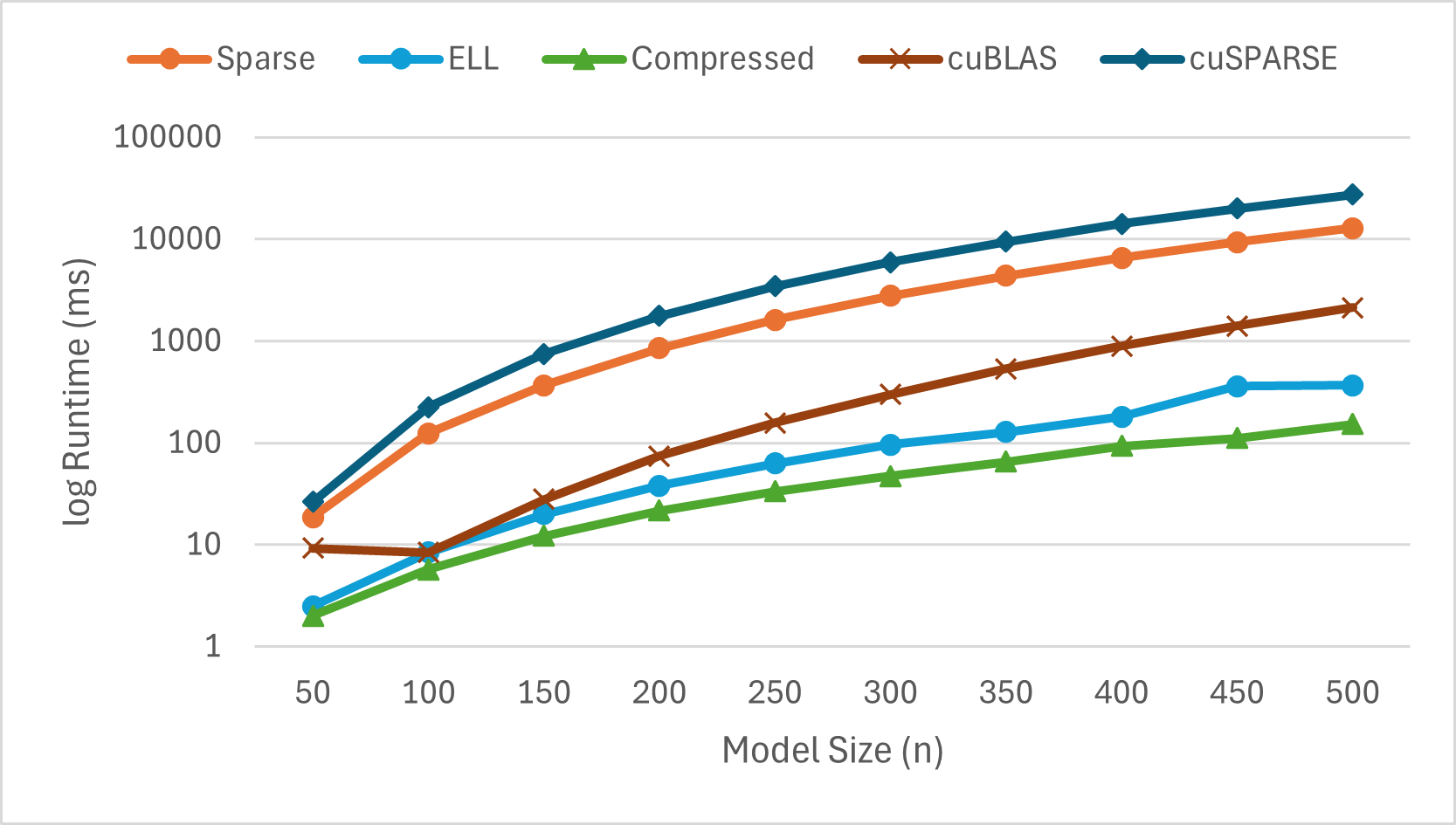}
		\caption{Execution time for the SNP systems sorting natural numbers on a RTX2080. X-axis shows the amount of natural numbers to sort (initially in descending order). Y-axis shows the time in ms using log scale.}
		\label{fig:sort_time_size}
\end{figure}

Figure \ref{fig:sort_time_size} shows the runtime of the different implementations over an RTX2080 with the SNP systems for sorting natural numbers (no delays). It is possible to observe that the most inefficient version of the simulator is the one using the cuSPARSE library. This may be due to the fact that the CSR format was required for this purpose, and it adds an important overhead since the transition matrix has to be compressed into that format. The Sparse version, whose transition matrix is not compressed, is the next slowest. For the cuBLAS version, a compression is not required, so it is possible to observe a better performance. ELL version follows in efficiency, and finally, Compressed version, which provides better overall performance. The maximum speedup is 83x, obtained for Compressed over Sparse for 500 numbers, while 34x is obtained for ELL over Sparse and 2.4x of Compressed over ELL. cuBLAS obtains 13x over Sparse for 150 numbers, but 6x for 500. However, cuSPARSE is always slower, around 0.4x.

\begin{figure}[H]
		\centering
		\includegraphics[width=\linewidth]{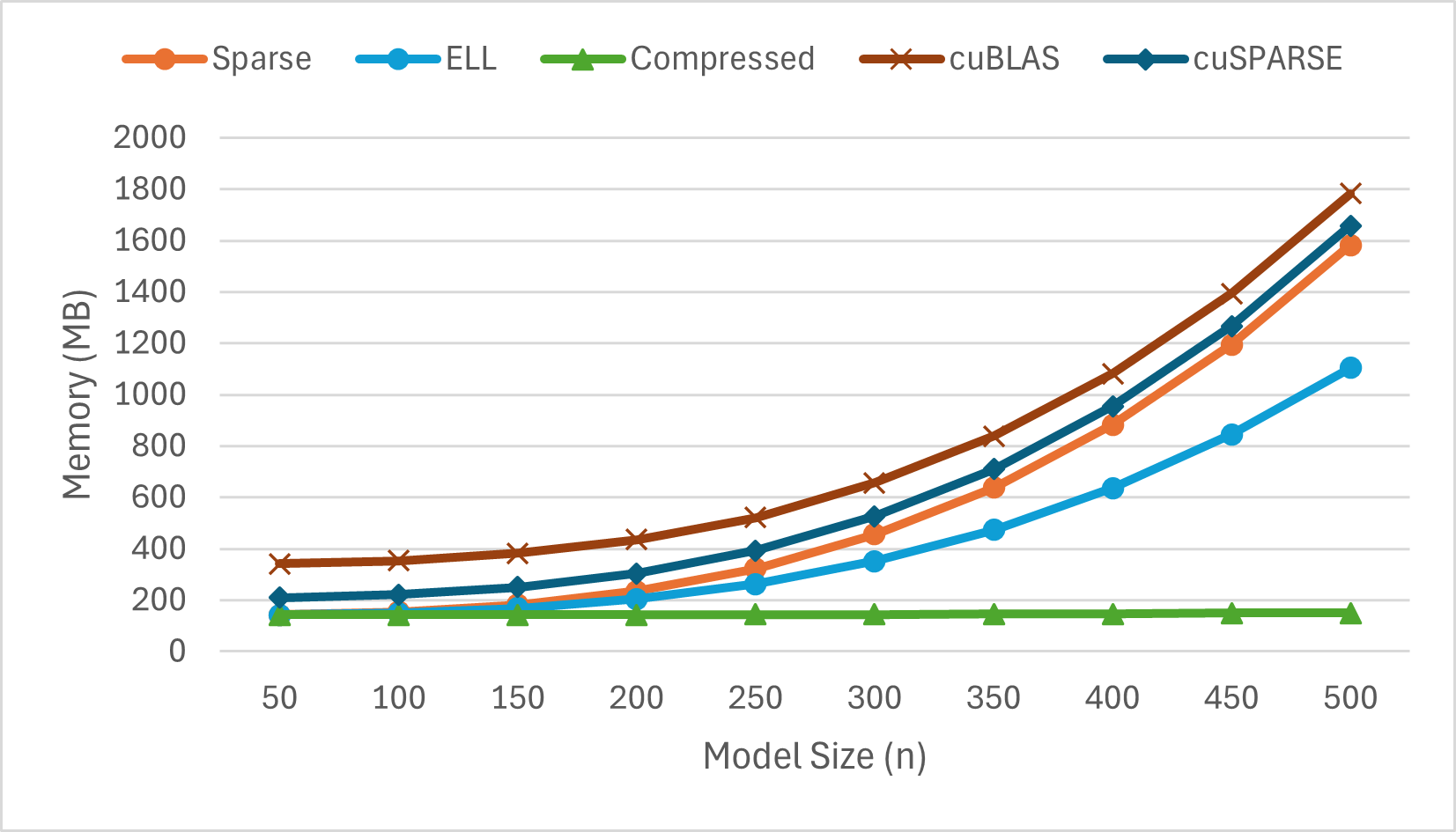}
		\caption{Memory consumption for the SNP systems sorting natural numbers. X-axis shows the amount of natural numbers to sort (initially in descending order). Y-axis shows the consumed memory in MB.}
		\label{fig:sort_mb_size}
\end{figure}

Figure \ref{fig:sort_mb_size} shows the corresponding amount of GPU memory required for the execution of each implementation. Once again the Compressed version is the lead, due to its efficient compression system. It is followed, in this order, by the ELL, Sparse, cuSPARSE and cuBLAS versions. Let us recall that cuSPARSE and cuBLAS require as a minimum data type to be 32bits (float), while our personalized compression methods use half integers. The Compressed version scales at a very low pace for this family of SNP systems, being almost constant up to 500 numbers. From our experiments, at 5000 the Compressed method requires 1GB, which is the case for ELL at 500 numbers. Overall, ELL saves 40\% of memory compared to Sparse, Compressed is 10 times smaller than Sparse and 7 times smaller than ELL. cuBLAS and cuSPARSE requires slightly more memory than Sparse, around 0.8 and 0.9 times, respectively.

\begin{tablehere}
\tbl{Total execution time of kernels (in ms) for computing spiking and configuration vector, for Sparse, ELL and Compressed formats, on a RTX2080. The employed model is an SNP system sorting 100 natural numbers. 
\label{table:kernels}	}
{\begin{tabular}{|c|c|c|c|}
	\hline 
	Kernel & Sparse & ELL & Compressed \\ 
	\hline
	\hline 
	Configuration vector & 1350.60 & 21.265 & 18.992 \\ 
	\hline 
	Spiking vector & 6.114 & 7.27 & 7.286 \\ 
	\hline 
\end{tabular} }
	%\centering
\end{tablehere}

Table \ref{table:kernels} shows the execution time of the two main CUDA kernels (both computing the spiking vector and the configuration vector at each transition) on the RTX2080. First of all, for this model, using compressed representations of matrices for simulating SNP systems is much better than using sparse representation. ELL is up to 63.5 times faster for configuration vector kernel, but Compressed is only 11\% faster than ELL. For spiking vector, the computation is a bit slower (17\%) in ELL and Compressed. Although the spiking vector is smaller in these versions (number of neurons instead of rules), the kernel is a bit affected.

A final experiment with this example is to run the Compressed method on the A100 80GB. The input size is increased gradually, so  the last configuration that fits the GPU is  $n=46,000$. It required $73,142$MB (71GB) of memory and took 6,923,465ms (1.9 hours) to complete the simulation. This instance, the largest to fit in 80GB of a high-end A100 GPU, accounts $198,000$ neurons and $2,116,046,000$ rules. This demonstrates that the proposed format enables the simulator to run larger instances.

\begin{figure}[H]
		\centering
		\includegraphics[width=\linewidth]{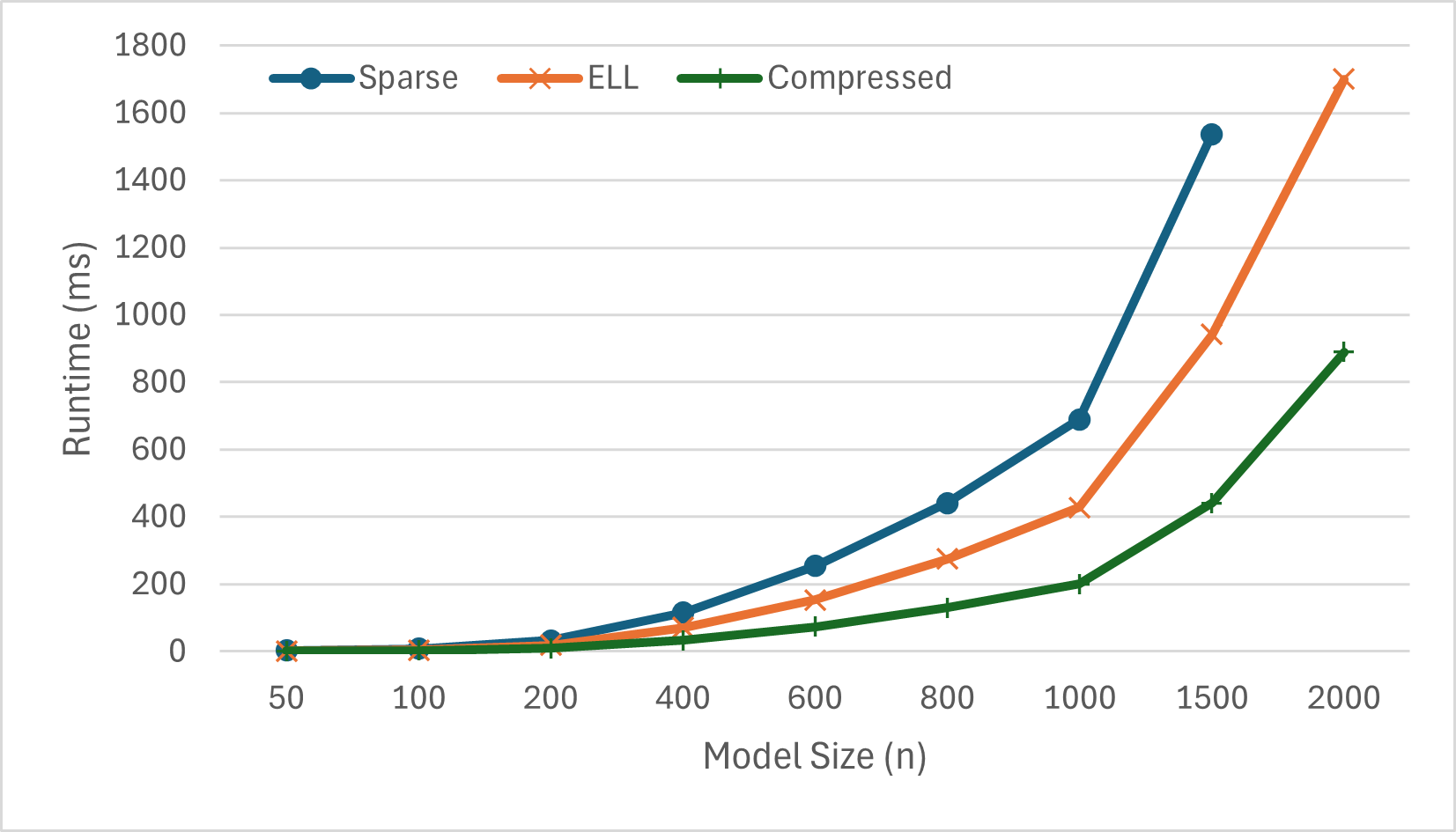}
		\caption{Execution time for the SNP systems solving  subset sum on a RTX2080. X-axis shows the size of the instance, measured as the size of the input set of numbers for the problem. The time is measured in ms.}
		\label{fig:subsetsum_time_size}
\end{figure}

\begin{figure}[H]
		\centering
		\includegraphics[width=\linewidth]{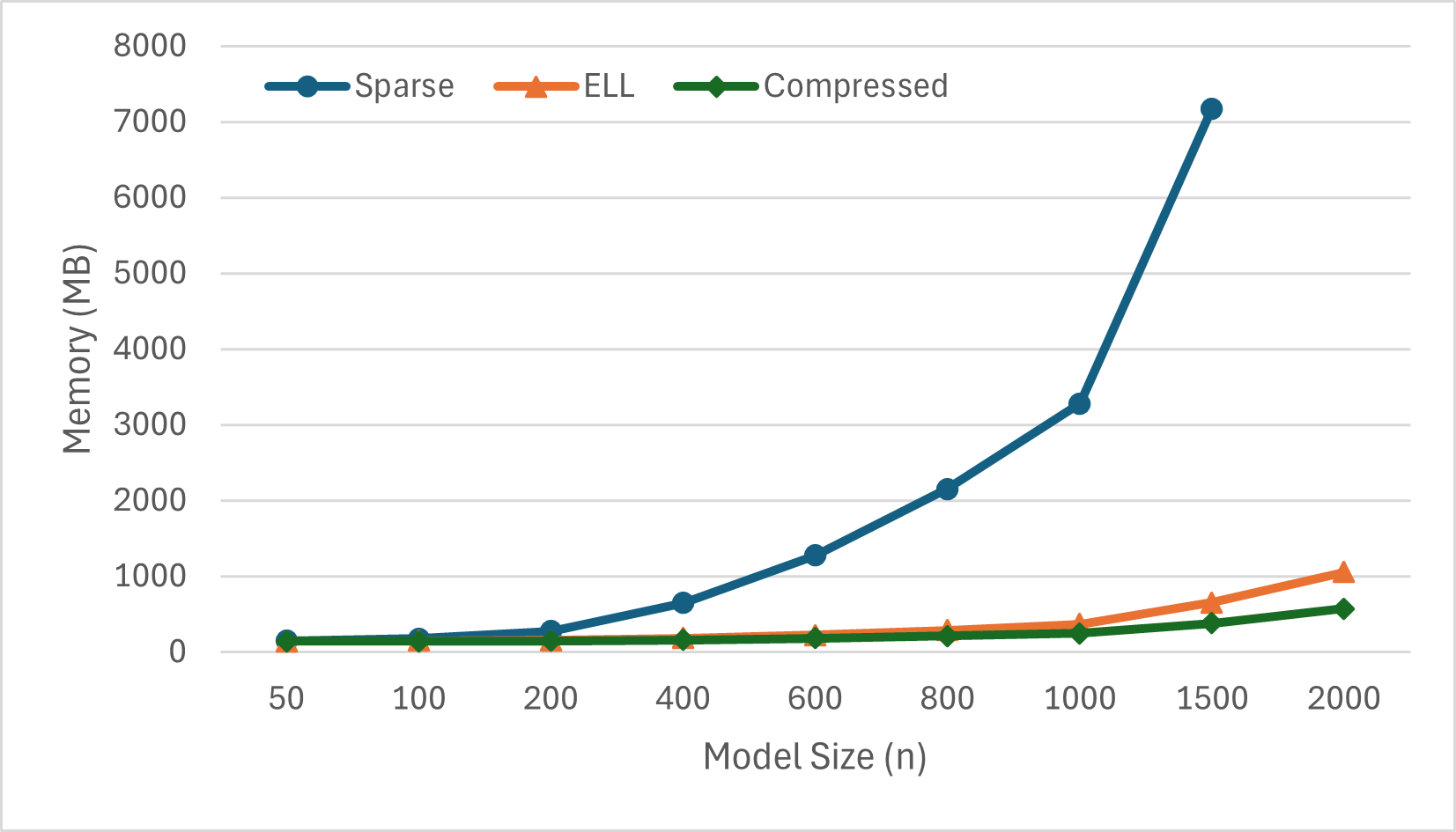}
		\caption{Consumed GPU memory for SNP systems solving  subset sum on a RTX2080. X-axis shows the size of the instance, with  memory  measured in MB.}
		\label{fig:subsetsum_mem_size}
\end{figure}

Figures \ref{fig:subsetsum_time_size} and \ref{fig:subsetsum_mem_size} show the execution time and the consumed memory for the SNP systems solving the subset sum problem. This solution uses delays, so our personalized versions (sparse, ELL and Compressed) can only be used, and not cuBLAS neither cuSPARSE. Moreover, this solution is non-deterministic, while our simulator is pure deterministic (i.e. it does not simulate non determinism). Therefore, the simulation runs only one computation path, so one execution is not enough to solve the provided subset sum instance. However, it is an approximation in order to test the performance (executed 3 times). The results are similar to those obtained in the previous example, reinforcing the observations. For instance, the Sparse version cannot handle an instance of size 2,000 on a RTX2080. For size 1,500, the obtained speedup of Compressed over Sparse is 3.5x and 2.1x over ELL, while ELL over Sparse is 1.63x. For memory footprint, ELL requires 11x less memory than Sparse, while Compressed requires 18.8 times less memory than Sparse and 70\% less than ELL.

\begin{figure}[H]
		\centering
		\includegraphics[width=\linewidth]{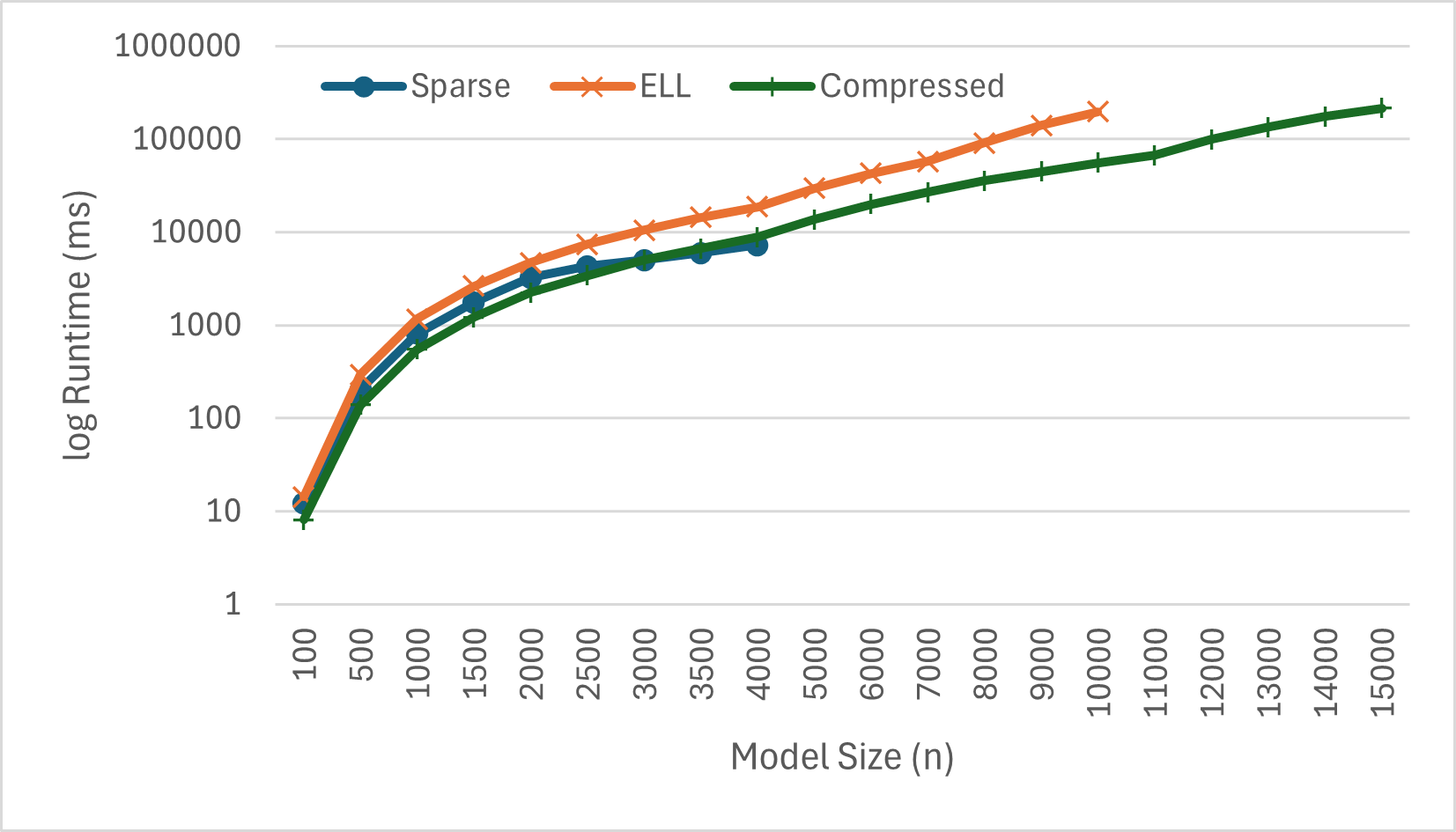}
		\caption{Execution time for the SNP systems solving  subset sum  on an A100 80Gb. X-axis shows the instance size, with time  measured in ms and shown at a log scale.}
		\label{fig:subsetsum_time_size_a100}
\end{figure}

\begin{figure}[H]
		\centering
		\includegraphics[width=\linewidth]{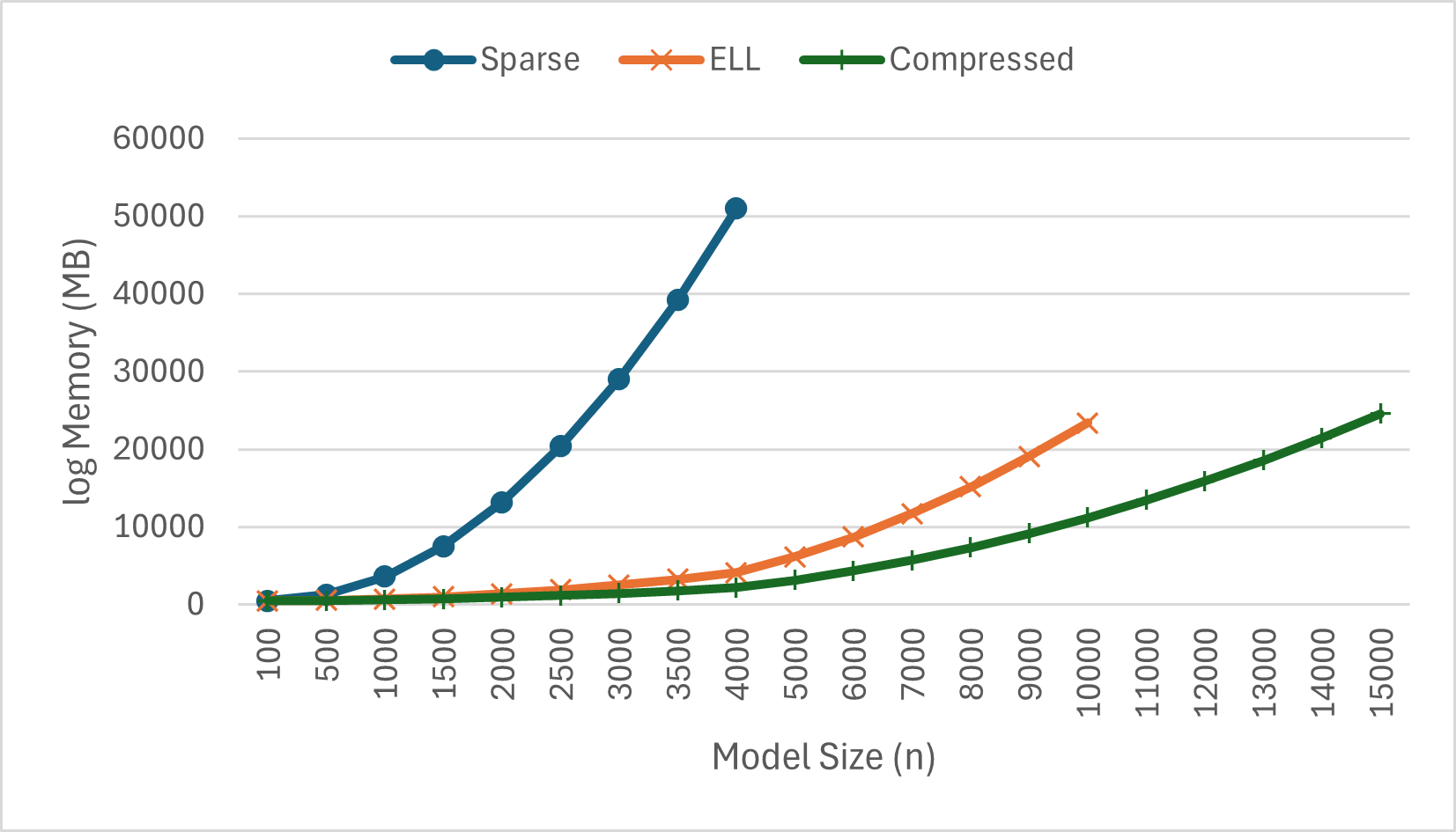}
		\caption{Consumed GPU memory for the SNP systems solving  subset sum  on an A100 Gb. X-axis shows the size of the instance. The memory is measured in MB.}
		\label{fig:subsetsum_mem_size_a100}
\end{figure}

Finally, Figures \ref{fig:subsetsum_time_size_a100} and \ref{fig:subsetsum_mem_size_a100} show the scalability of the simulator on the subset sum problem, when running on an A100 with 80Gb (one of the GPUs with the largest amount of memory in the market to date of writing). The sparse version cannot handle an input size of 5,000 on that GPU, since for 4,000 it requires around 50Gb of memory. It is interesting to see that the runtime of the sparse version is very similar to the Compressed one. ELL version scales better but restrictions on how to construct the initial matrix on CPU makes the simulator not able to handle an input size of 11,000. On the contrary, Compressed version can go to input sizes of 15,000 and beyond.

\section{Conclusions and Future Work}
\label{sec:conclusions}

In this paper, GPU implementations of compressed matrix representations for SNP systems are introduced. The first version is  tested with a family of SNP systems sorting natural numbers in parallel, and a solution to the {\bf NP}-complete problem Subset Sum. It is reported up to 83$\times$ of speedup of the Compressed method compared to sparse representation, and 34$\times$ of ELL against Sparse (for Sort example on 500 numbers). The proposed formats not only enable speedups, but also to scale to larger instances, as tested with an A100 80GB. For the Subset Sum example, up to 3.5$\times$ of speedup is reported. Although both examples have a maximum out degree in the neurons ($z$) equal to the instance size ($n$), the compression ratio is slightly higher for Subset sum problem according to the experiments. However, the achieved speedup  is higher for the Sorting problem. This fact indicates that Subset sum example is less sparse than the Sorting one:  the amount of work remaining in compressed matrices is less accelerated on GPUs, plus requiring to handle with delays.

The results can be further improved. The future plan is to optimize the kernels, putting more efforts on improving the parallelism to better fit the GPU architecture, such as the new sparse capabilities in Tensor Cores. Other examples are being tested with our simulators to better characterize them, such as sorting networks\cite{rodica_sortsnp08}. Moreover, the simulators are being extended to  support SNP systems with dynamic structures (budding, division and plasticity) and supporting non-deterministic solutions. For instance, it was shown\cite{processes} that the idea of synapse plasticity (adding or removing synapses alone) can be better suited for GPU implementation, compared to adding or removing both neurons and synapses\cite{gatti_snpneursyn2022}.   

Furthermore, we aim to handle a more general type of regular expression, denoted as {\it type four (4)},  of the form  $a^i(a^j)^*$ where $j \geq 0$ and $i \geq 1$, and $*$ can be replaced with $+$. Later, perhaps type 4 expressions are enough for simulations, instead of supporting three types of expressions in Section \ref{sec:baserep}. Recently it was shown\cite{ivan22_normsnp}  that type 4 expressions are enough for SNP systems to maintain Turing completeness. The present work can also be used to improve the automatic design of SNP systems\cite{lovely_esnp21}, especially on GPUs\cite{evolsnpgpu22}. More types of SNP systems can be also considered for this representation, starting by improving the representation for SN P systems with plasticity\cite{zak_snpsnpmat2019}, other variants mentioned in recent surveys\cite{snpsurv_naco2022,francis_acmc2022jour}, including recent results on some variants.\cite{wu2022tuning}.

The main target is also to provide a flexible {\it framework} to simulate SNP systems, by providing an API in common languages such as Python or C++, in order to programmatically define and simulate SNP systems. This is inspired from modern Deep Learning frameworks such as Keras and PyTorch. P-Lingua 5 will be also employed in order to define the SNP models from text files with a syntax close to the one used by model designers.
In this way, our target framework can be used as a main component for real world applications with SNP systems and variants, such as edge detection\cite{edge_det2023}, data forecasting\cite{forecast_snp2023,nsnp_forecast2022}, classification with supervised learning\cite{layered_snp2022}, cryptosystems\cite{snp_crypto2022}, robots\cite{esnp_robot2022}, sentiment analysis\cite{lst_snpsenti2023}.
The work presented in this paper offers a promising impact on both runtime and memory improvements on such real world problems.
Together with the development of our target framework we also aim to include an interactive and visual component\cite{websnapse_page,ravsim2023,gulapa2023reloaded}: in this way users can easily perform experiments without having deep knowledge of GPUs and matrix representations. 

The experiments here presented focus mainly on CUDA GPUs. Such GPUs currently provide the best experience to researchers in terms of programming and technical support for massively scalable and parallel computing. Despite the focus of the experiments in CUDA GPUs, this work should still be applicable to other SPMD or SIMD GPUs or processors. Therefore, in order to reach a larger amount of users, it is also proposed to consider open GPU standards like OpenCL and SYCL, as well as ROCm for AMD GPUs. 
For instance, a preliminary work exists\cite{snp-ocl-16pcsc} on an OpenCL simulator.

As a general remark, it can be concluded that spiking neural P system model is convenient for parallelization, since deciding whether a neuron spikes (and by which rule) does not depend on other neurons, while updating a configuration can be realized by scatter operations (e.g. atomic operations on GPUs) \cite{dend-tool}. Nevertheless, synchronization is necessary between the transition steps, and between choosing the spiking vector and updating the configuration.
Detailed comparisons in  the theory and practice of SN P systems and SNNs\cite{ghosh_snn2009} is of great interest, such as training\cite{ghosh_snn2007} and learning\cite{ghosh_mulsnn2009} algorithms.

\section*{Acknowledgements}

The A100 80GB GPU was obtained with projects EQC2019-006325-P of the Spanish ``Ministerio de Ciencia, Innovaci\'on y Universidades'' with FEDER funds, and  IE10\_118 USE of the Andalusian ``Consejer\'ia de Econom\'ia, Conocimiento, Empresas y Universidad'' with FEDER funds. An A100 40GB, donated by the NVIDIA Hardware Grant, was also tested.
This work is also supported by the Zhejiang Lab BioBit Program (Grant No. 2022BCF05).
F.G.C. Cabarle is supported by the \textit{QUAL21 008 USE} project (PAIDI 2020 and FEDER 2014-2020 funds). 

\bibliographystyle{ws-ijns}
\bibliography{mybibliography}
\end{multicols}
\end{document}